\pdfoutput=1
\documentclass[aip,rsi,groupedaddress,reprint,showpacs,nofootinbib,floatfix,longbibliography,showkey,samsmath,amssymb,]{revtex4-2}
\usepackage[USenglish]{babel}
\usepackage[utf8]{inputenc}
\usepackage{graphicx}
\usepackage{amsmath} 
\usepackage{amsfonts}
\usepackage{hyperref}
\hypersetup{
	pdfstartview={FitH},
	pdftitle={Mathematical analysis of the operation of a scanning Kelvin probe},
	pdfauthor={Edgar A. Pina-Galan},
	pdfsubject={Kelvin probe},
	pdfcreator={Edgar A. Pina-Galan},
	pdfproducer={Edgar A. Pina-Galan},
	pdfkeywords={scanning Kelvin probe, contact electrification, contact potential difference, work function},
	colorlinks,
	citecolor=black,
	filecolor=black,
	linkcolor=black,
	urlcolor=black}
\makeatletter
\let\frontmatter@footnote@produce\frontmatter@footnote@produce@endnote
\makeatother
\begin{document}
\title{Mathematical analysis of the operation of a scanning Kelvin probe}
\author{\firstname{Edgar A.} \surname{Pina-Galan}}
\email{edgar.pinagalan@gmail.com}
\affiliation{Manchester Institute of Biotechnology, University of Manchester, Manchester M13 9PL, United Kingdom}
\altaffiliation{Current affiliation: Shenzhen Engineering Laboratory for Precision Medicine and Healthcare, Tsinghua-Berkeley Shenzhen Institute, Tsinghua University, Shenzhen 518055, People’s Republic of China}
\date{\today}
\newcommand\blfootnote[1]{%
	\begingroup
	\renewcommand\thefootnote{}\footnote{#1}%
	\addtocounter{footnote}{-1}%
	\endgroup
}
\begin{abstract}
The scanning Kelvin probe is a tool that allows for the contactless evaluation of contact potential differences in a range of materials, permitting the indirect determination of surface properties such as work function or Fermi levels. In this paper, we derive the equations governing the operation of a Kelvin probe and describe the implementation of the off-null method for contact potential difference determination, we conclude with a short discussion on design considerations.
\end{abstract}
\maketitle
\section{Introduction}
\blfootnote{Copyright (2019) Edgar Alexis Pina Galan. This article is distributed under a Creative Commons Attribution (CC BY) License.}

The Kelvin probe is a non-contact, non-destructive technique based on the so-called contact electrification that occurs in a parallel-plate capacitor when electrical contact between its plates is established. \cite{Kelvin1898} This tool is used to evaluate information on a surface topography;\cite{Zisman1932,Mackel1993} typically used to measure work function in metals,\cite{Fernandez-Garrillo2018,Kotani1971} it can also quantify surface potential difference for non-metals \cite{Baikie1999,Ahn2012,Baikie1998,Kotani1971} or Fermi levels in organic thin films. \cite{Pfeiffer1996}

The principle of operation was first proposed by Kelvin in 1898 as a means to measure the contact potential difference $V_\textrm{CPD}$ of two electrically connected metals without physical contact between each other inner surfaces.\cite{Kelvin1898} The method was based on the measurement of the surface charge variation $\Delta Q_s$ between the sample surface and an electrically conductive reference electrode in a parallel-plate capacitor configuration in order to determine the $V_\textrm{CPD}$ between both materials. A backing potential $V_\mathrm{b}$ was applied to one of the plates, then, displacing the reference electrode a certain distance $\Delta d$ from the sample surface will induce a change in capacity $\Delta C$, Kelvin then used a galvanometer to measure the surface charge $Q_s$. This process should be repeated until $\Delta Q_s=0$ which indicated that the potential difference between the sample material and reference electrode was equal to the bias voltage (\textit{i.e.}, $-V_\textrm{CPD}=V_\mathrm{b}$).

Zisman improved this experiment in 1932 by using a piano wire to set the reference electrode into vibration, generating an oscillating capacitive charge $C_K(t)$ between the two surfaces.\cite{Zisman1932} The result is an electric current $i(t)=\delta Q_s(t)/\delta t$ induced on the reference electrode which was further converted into an audio-frequency signal. A backing potential $V_\mathrm{b}$ was applied and manually adjusted until no signal was heard, indicating that $V_\mathrm{b}$ equalled $-V_\textrm{CPD}$.

Later, in 1999, Baikie presented a Kelvin probe system fully controlled by a computer.\cite{Baikie1999} The inclusion of a computer allowed for the introduction of the so-called off-null method for determining $V_\textrm{CPD}$.  Additionally, by maintaining a constant probe-to-sample distance, it became possible to automatically scan a relatively irregular surface.

In this paper we provide the background and derive the equations governing the operation of a Kelvin probe and the off-null method for determining $V_\textrm{CPD}$. The equations provided can be readily applied to the simulation or design of a scanning Kelvin probe system capable of obtaining reliable measurements.

\section{Analysis of the operation of a Kelvin probe}

\subsection{Operation principle}

\begin{figure}[h!]
	\centering
	\includegraphics[width=1\linewidth]{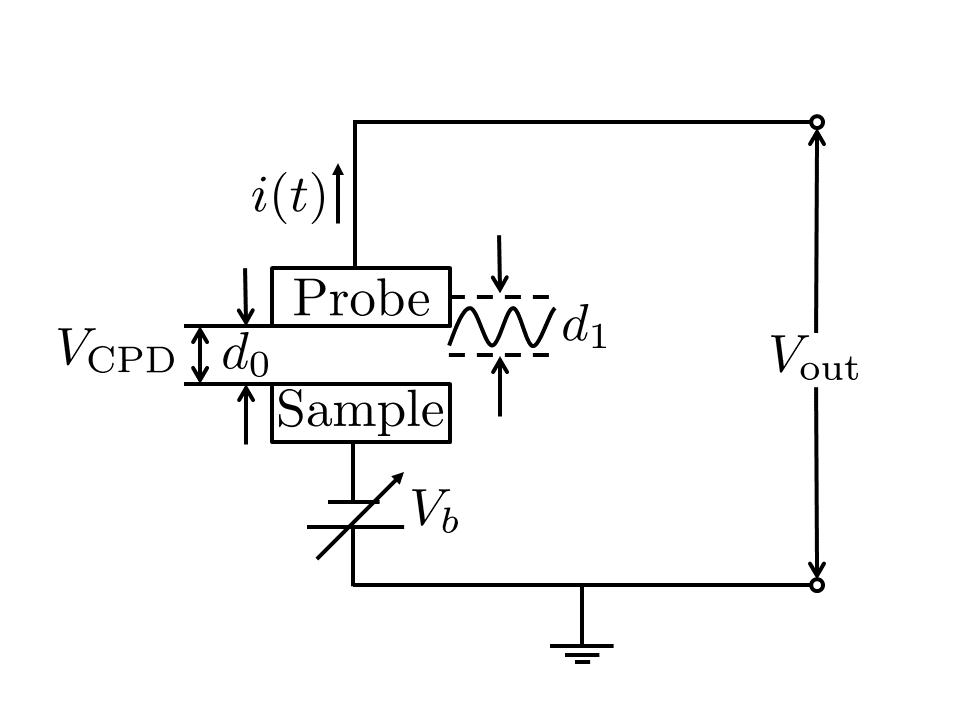}
	\caption{\label{fig:KPdiagram}Schematic of the principle of operation of a Kelvin probe. Where $V_\textrm{CPD}$ is the probe-to-sample contact potential difference, $d_0$ is the mean probe-to-sample distance, $d_1$ is the probe oscillation amplitude, $i(t)$ is the current induced on the probe tip, $V_\textrm{out}$ is the voltage output of the probe circuit, and $V_\mathrm{b}$ is the backing potential. Since the capacitance is a function of the separation between the plates, the vibration of the probe results in an oscillating capacitive charge, generating a current $i(t)$ that can be converted into a voltage $V_\textrm{out}$ for further processing of the signal.}
	
\end{figure}

The scanning Kelvin probe operates by creating a parallel-plate capacitor between an electrically conductive probe vibrating perpendicularly and the nearest surface it approaches (Fig. \ref{fig:KPdiagram}). When external electrical contact between probe and sample takes place by means of the application of a backing potential $V_\mathrm{b}$, the Fermi levels of the probe and sample surfaces start to equalize, resulting in a charge flow which in turn generates a contact potential difference $V_\textrm{CPD}$. Determination of $V_\textrm{CPD}$ permits the indirect calculation of surface properties such as work function, \cite{Fernandez-Garrillo2018} surface potential, \cite{Baikie1999,Ahn2012,Kotani1971} or Fermi levels. \cite{Pfeiffer1996}

\subsection{Contact potential difference}

The contact potential difference $V_{CPD}$ between the surfaces of two materials arises from a phenomenon referred to as contact electrification. Whenever two dissimilar materials are brought into contact, thermodynamic equilibrium mechanisms cause the electrochemical potentials of their charged particles to equalize. Therefore, the contact potential difference $V_{CPD}$ is related to the Fermi level and work function of the material.

We can define the work function of a surface as the difference in potential energy of an electron between the vacuum level and the Fermi level; \cite{Kittel1996} with a magnitude given by

\begin{equation}
\Phi=-e\phi-\epsilon_F,
\end{equation}

where $\Phi$ is the work function of the surface, $-e$ is the elementary charge, $\phi$ is the electrostatic potential in the vacuum, and $\epsilon_F$ is the Fermi level of the material. Therefore, the term $-e\phi$ represents the energy of an electron at rest in the vacuum.

\begin{figure}[h!]
	\centering
	\includegraphics[width=1\linewidth]{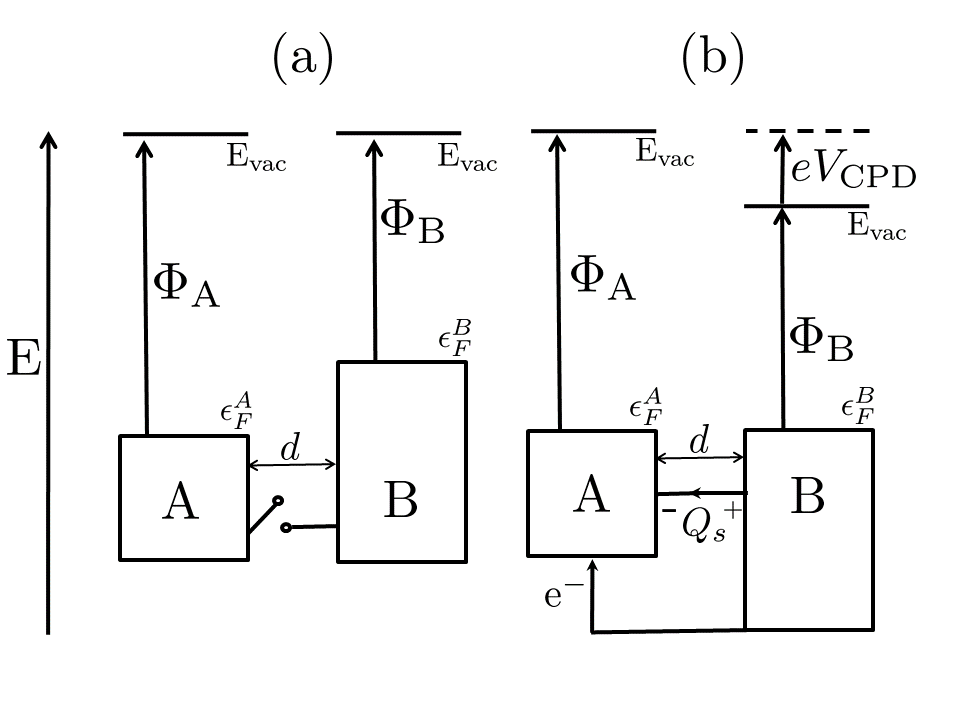}
	\caption{\label{fig:ContactElectrification}Electron energy level diagrams of two dissimilar materials, $\textrm{A}$ and $\textrm{B}$, separated by a distance $d$: (a) without contact and (b) with external electrical contact, allowing a surface charge flow $Q_s$ that originates a contact potential difference $V_{CPD}$. $\epsilon_F^A$ and $\epsilon_F^B$ refer to the Fermi levels, and $\Phi_A$ and $\Phi_B $ are the work functions of the materials $\textrm{A}$ and $\textrm{B}$, respectively.}	
	
\end{figure}

With this in mind, a contact electrification scenario can then be depicted by the system in Fig. \ref{fig:ContactElectrification}. When two dissimilar materials, $\textrm{A}$ and $\textrm{B}$, are approached with a distance $d$ separating their inner surfaces, and considering that $\Phi_A\neq\Phi_B$: (a) if there is no physical contact between materials $\textrm{A}$ and $\textrm{B}$, electrons remain at their respective Fermi levels, $\epsilon_F^A$ and $\epsilon_F^B$, with work functions $\Phi_A$ and $\Phi_B$, respectively; (b) when electrical contact is established, highest energy electrons on the outermost level bands will flow from the lowest- to the highest-work function material until all particles at the Fermi levels $\epsilon_F^A$ and $\epsilon_F^B$ are distributed uniformly and equilibrium is reached. Work function remains unchanged as it being an intrinsic property of materials. For the case where $\Phi_A>\Phi_B$, electrons flow from surface $\textrm{B}$ to $\textrm{A}$, this flux polarizes the inner faces of the materials with the negative pole at the surface receiving the electrons, resulting in a contact potential difference $V_\textrm{CPD}$ equivalent to the difference in work functions of the materials $\textrm{A}$ and $\textrm{B}$; this is expressed as

\begin{equation}
V_\textrm{CPD}=\frac{(\Phi_B-\Phi_A)}{e}=\frac{(\epsilon_F^A-\epsilon_F^B)}{e}.
\label{CPD}
\end{equation}

The contact potential difference $V_{CPD}$ can then be defined as the energy required to transfer an electron from one surface to another between two materials with dissimilar work functions. Since $V_{CPD}$ is related to $\Phi$ and $\epsilon_{F}$, it is possible to indirectly determine $\Phi$ or $\epsilon_{F}$ when measuring $V_{CPD}$ between a sample and a reference material with known $\Phi$ or $\epsilon_{F}$, respectively.

In order to determine the contact potential difference $V_\textrm{CPD}$, a known voltage can be applied between the probe and sample and varied until there is no measurable current induced on the probe, indicating that its magnitude has equalled $V_\textrm{CPD}$, this traditional concept is referred to as the null method. However, in modern devices, a so-called off-null approach is rather employed due to the higher signal-to-noise ratio provided as well as allowing the automatic scanning of an irregular surface. 

\subsection{\label{offnull}Off-null method}

The off-null method can be employed in order to determine the probe-to-sample contact potential difference $V_\mathrm{CPD}$ as well as to maintain a constant mean probe-to-sample distance $d_0$ during automatic scanning with a Kelvin probe. Maintaining a constant $d_0$ is essential to prevent erroneous $V_\mathrm{CPD}$ readings arising from spacing changes. Its implementation involves lower costs when compared with the traditional null method, \cite{Baikie1998} which requires the use of a lock-in amplifier (widely employed in atomic force microscopes). Additionally, it offers a higher signal-to-noise ratio by avoiding common sources of measurement errors related to the null point such as stray capacitance or talk-over noise from the actuator. \cite{Baikie1991a,Baikie1991b} 

\begin{figure}[h!]
	\centering
	\includegraphics[width=1\linewidth]{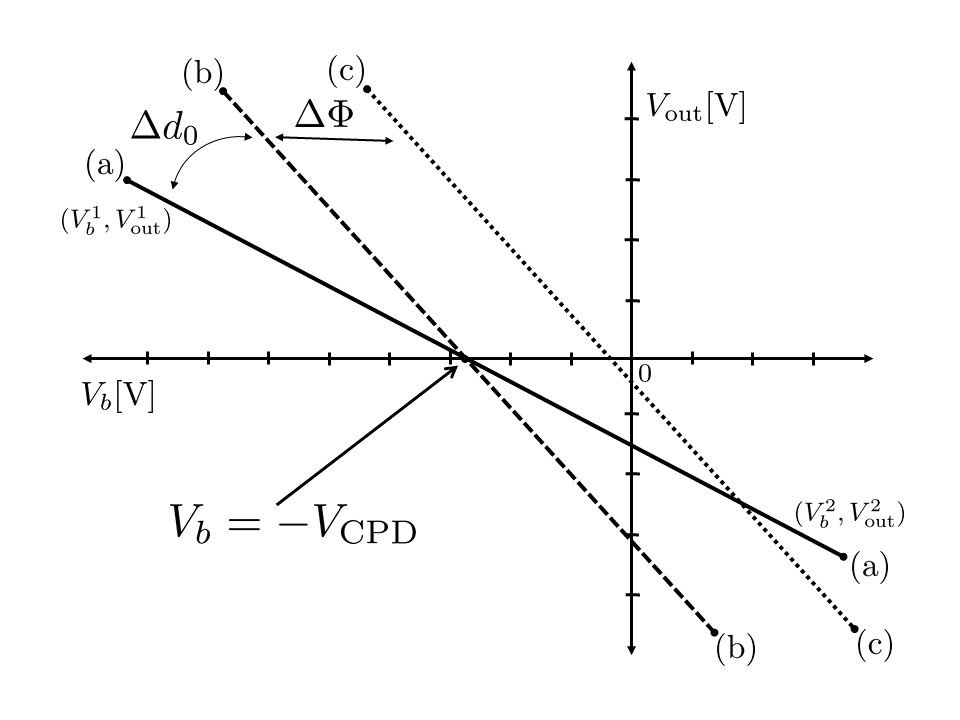}
	\caption{\label{fig:baikie1}Schematic diagram representing the output voltage $V_\mathrm{out}$ vs backing potential $V_\mathrm{b}$ function: curve (a) represents the plotted function of the form $V_\mathrm{out}=m_K V_\mathrm{b}+c$, (b) shows a change in the gradient $m_K$ induced by a mean probe-to-sample distance $d_0$ variation, and (c) shows the effect of a variation in work function $\Phi$ on the sample material.}
	
\end{figure}

The method consists on applying at least two values of a backing potential $V_\mathrm{b}$, the value for $V_\textrm{CPD}$ can subsequently be retrieved from the linear relation between $V_\mathrm{b}$ and the voltage $V_\textrm{out}$ delivered by the device, which is a linear function of the form

\begin{equation}
V_\textrm{out}=m_K V_\mathrm{b}+c,
\end{equation}

where the value for $V_\mathrm{b}$ corresponding to $V_\textrm{out}=0$ is equal to $V_\textrm{CPD}$, \textit{i.e.}, $V_\mathrm{b}=-V_\textrm{CPD}$, and $m_K$ is termed the Kelvin gradient (Fig. \ref{fig:baikie1}).

Therefore, given the two backing potential values $V_\mathrm{b}^1$ and $V_\mathrm{b}^2$, that elicit the responses $V_\mathrm{out}^1$ and $V_\mathrm{out}^2$, respectively, and assuming that the value of $V_\mathrm{CPD}$ lies anywhere between $V_\mathrm{b}^1$ and $V_\mathrm{b}^2$, $V_\mathrm{CPD}$ is calculated by simple linear interpolation as

\begin{equation}
V_\mathrm{CPD}=V_\mathrm{b}^1+(V_\mathrm{b}^2-V_\mathrm{b}^1)\frac{-V_\mathrm{out}^1}{V_\mathrm{out}^2-V_\mathrm{out}^1}.
\end{equation}

Additionally the Kelvin gradient $m_K$ of this function is inversely proportional to the mean probe-to-sample distance squared (\textit{i.e.}, $m_K\propto d_0^{-2}$); this will be deduced in section \ref{circuit}. Hence, a constant value for $m_K$ can be maintained during scanning by calculating its value as 

\begin{equation}
m_K=\frac{V_\mathrm{out}^2-V_\mathrm{out}^1}{V_\mathrm{b}^2-V_\mathrm{b}^1}
\end{equation}

and instructing a translation system to compensate any variations with a corresponding adjust of $d_0$ \textit{via} a feedback loop. Maintaining a constant mean probe-to-sample distance $d_0$ is an essential requirement to obtain reliable measurements over an irregular surface. \cite{Ritty1981}

\section{\label{circuit}Kelvin Probe Electric Circuit Analysis}

\begin{figure}[h!]
	\centering
	\includegraphics[width=1\linewidth]{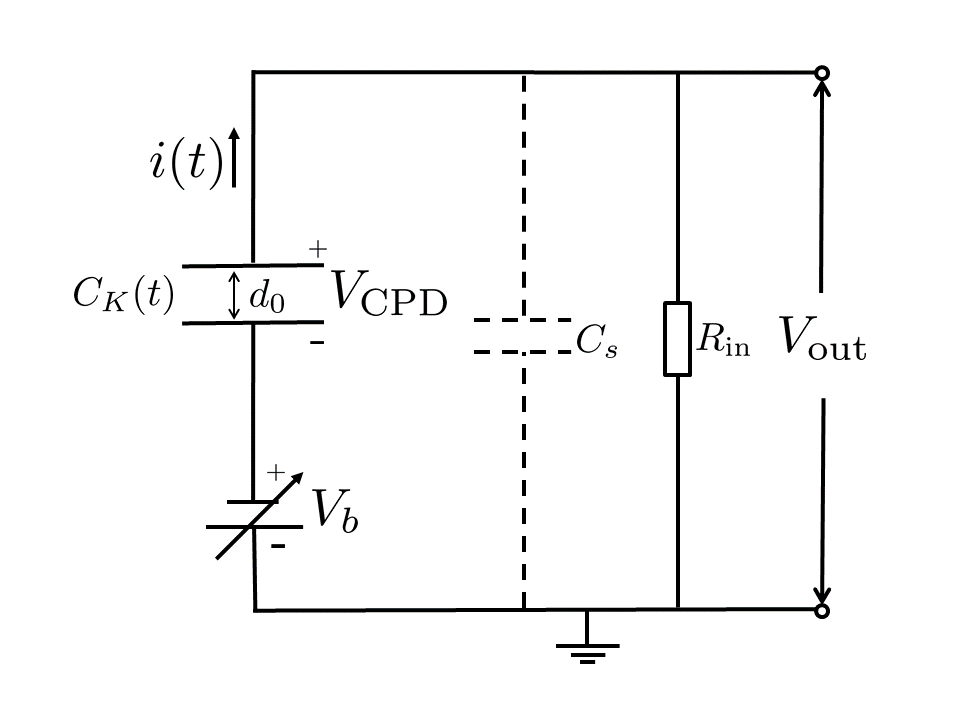}
	\caption{\label{fig:kpcircuitreduced}Reduced circuit of a Kelvin probe, where the $I/V$ converter is represented by its effective resistance $R_\textrm{in}$ and current lost due to stray capacitance $C_s$ is neglected.}
	
\end{figure}

Let us consider the electric circuit in Fig. \ref{fig:kpcircuitreduced}, which represents a parallel-plate capacitor formed between a sample surface and an electrically conductive reference electrode, where $V_\textrm{CPD}$ is the contact potential difference generated between the plates, $d_0$ is the mean probe-to-sample distance, $C_K(t)$ is the time-varying capacitance, and $i(t)$ is the current induced due to the oscillation of the reference electrode. A backing potential $V_\mathrm{b}$ is introduced to close the circuit, while $V_\textrm{out}$ is measured in order to determine $V_\textrm{CPD}$. The Kelvin probe usually consists of an operational amplifier as a current-to-voltage $I/V$ converter component whose output is further amplified and filtered before being input to a data acquisition system. For this analysis, the Kelvin probe circuit has been reduced by replacing the $I/V$ converter for its effective resistance $R_\textrm{in}$. Additionally, current lost due to stray capacitance $C_s$ is neglected under the assumption that the contribution of $C_s$ in the magnitude of the effective impedance of the circuit is minimal. \cite{Baikie1999} An analysis of stray capacitance on the Kelvin probe circuit can be found in Refs. \onlinecite{Baikie1991a,Hadjadj1995}.

According to Gauss's law, the electric field between the plates of a parallel-plate capacitor is given by 

\begin{equation}
\overrightarrow{E}=\frac{Q_s}{\varepsilon_r\varepsilon_0 A_p},
\end{equation}

which for our purpose can be rewritten as 

\begin{equation}
\overrightarrow{E}d_0=V_\textrm{CPD}=\frac{Q_s d_0}{\varepsilon_r\varepsilon_0 A_p},
\label{Gauss}
\end{equation}

where $\overrightarrow{E}$ is the electric field magnitude, $Q_s$ is the surface charge, $A_p$ is the area of the probe interacting with the sample surface, and $\varepsilon_r$ and $\varepsilon_0$ are the relative permittivity of the medium and vacuum, respectively. 

Since charge is defined as

\begin{equation}
Q_s=V_\textrm{CPD}C_K,
\label{Charge}
\end{equation}

it can be noted from Eqs. \ref{Gauss} and \ref{Charge} that

\begin{equation}
C_K=\frac{\varepsilon_r\varepsilon_0 A_p}{d_0}.
\label{Ck}
\end{equation}

For a parallel-plate capacitor with an average gap $d_0$ between its plates, when a sinusoidal displacement with amplitude $d_1$ and frequency $f$ is introduced on one of the plates, the distance $d(t)$ separating the plates over time is described by

\begin{equation}
d(t)=d_0+d_1\cos 2\pi ft=d_0(1+\gamma\cos\varphi),
\end{equation}

where $\gamma\equiv d_1/d_0$ is termed the modulation index, and $\varphi=2\pi ft$. Equation \ref{Ck} then becomes

\begin{equation}
C_K(t)=\frac{\varepsilon_r\varepsilon_0 A_p}{d_0(1+\gamma\cos\varphi)},
\label{C(t)}
\end{equation}

where $C_K$ is now referred to as the time-varying $C_K(t)$.

When a backing potential $V_\mathrm{b}$ is introduced, Eq. \ref{Charge} takes the form

\begin{equation}
\Delta Q_s=(V_\textrm{CPD}+V_\mathrm{b})\Delta C_K.
\label{Change in charge}
\end{equation}

If the probe is set into vibration, the oscillation of $C_K$ generates a current $i(t)$ on the probe due to the periodic change in charge $\delta Q_s(t)/\delta t$. Equation (\ref{Change in charge}) can then be rewritten as

\begin{equation}
\frac{\delta Q_s(t)}{\delta t}=i(t)=(V_\textrm{CPD}+V_\mathrm{b})\frac{\delta C_K(t)}{\delta t}.
\label{i(t)}
\end{equation} 

Equation \ref{C(t)} is now substituted into Eq. \ref{i(t)} to obtain

\begin{equation}
i(t)=(V_\mathrm{CPD}+V_\mathrm{b})\frac{\varepsilon_r\varepsilon_0 A_p}{d_0}\frac{\delta }{\delta t}\left[\frac{1}{(1+\gamma\cos\varphi)}\right].
\end{equation}

The derivative of $i(t)$ yields a function that describes the current induced at the vibrating probe tip as

\begin{equation}
i=(V_\mathrm{CPD}+V_\mathrm{b}){\left[\frac{\varepsilon_r\varepsilon_0 A_p}{d_0}\right]}{\left[\frac{2\pi f\gamma\sin\varphi}{(1+\gamma\cos\varphi){^2}}\right]}.
\label{current}
\end{equation}

The output voltage is then given according to Ohm's law by

\begin{equation}
V_\mathrm{out}=GR_f C_0\omega\gamma(V_\mathrm{CPD}+V_\mathrm{b})\left[\frac{\sin(\varphi+\alpha)}{(1+\gamma\cos(\varphi+\alpha)){^2}}\right],
\label{Vout}
\end{equation}

where $G=-R_2/R_1$ is the pre-amplifier gain of the Kelvin probe circuit, $R_f$ is the feedback resistance of the $I/V$ converter, $\omega=2\pi f$ is the angular vibration frequency, $\alpha$ is a phase angle, and 

\begin{equation}
C_0\equiv\frac{\varepsilon_r\varepsilon_0 A_p}{d_0}
\end{equation}

is termed the mean capacitance. 

\begin{figure}[h!]
	\centering
	\includegraphics[width=1\linewidth]{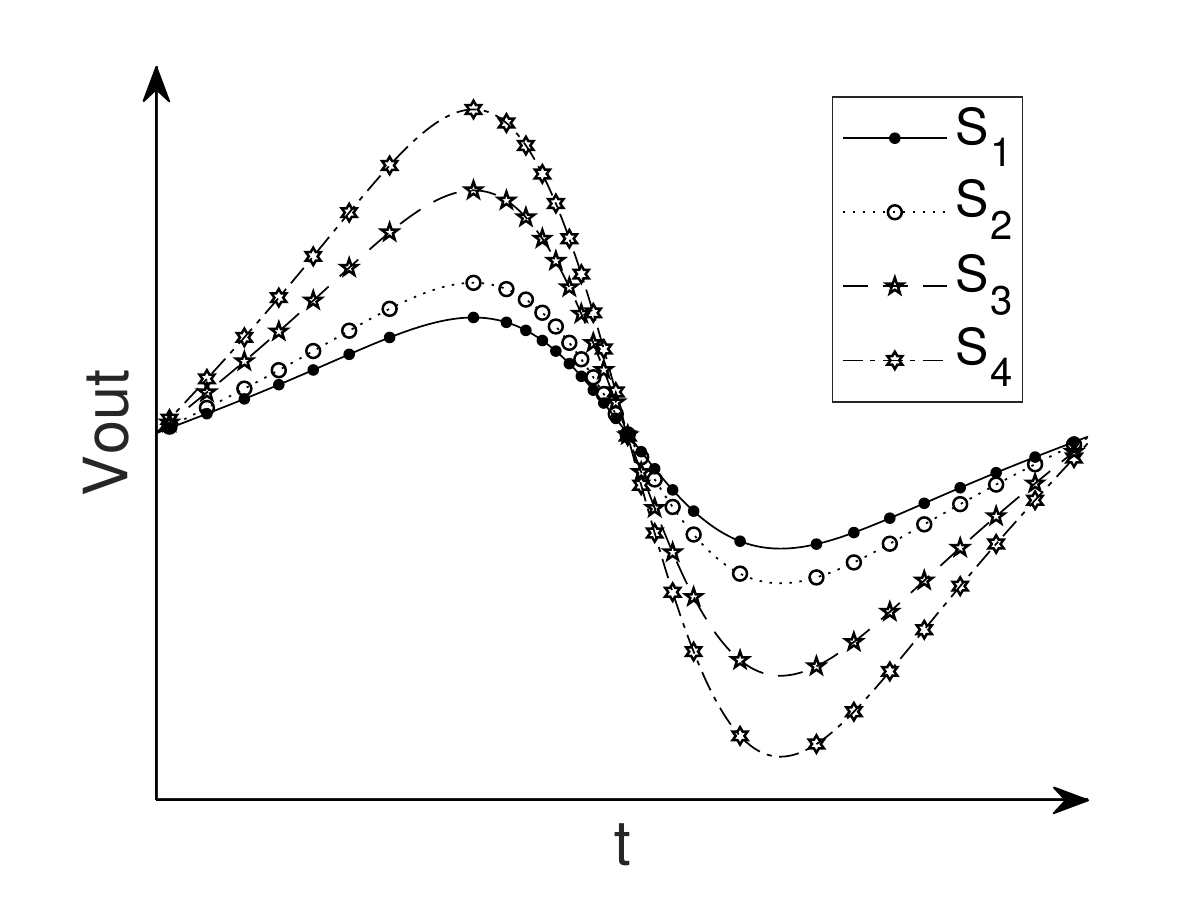}
	\caption{\label{fig:sensitivity}Representation of the effect of the sensitivity $S$ on a typical output signal $V_\mathrm{out}$ of a Kelvin probe circuit. Four arbitrary $S$ values are employed, where $S_1<S_2<S_3<S_4$.}
	
\end{figure}

The sensitivity $S$ directly affects $V_\mathrm{out}$ intensity on a Kelvin probe circuit (Fig. \ref{fig:sensitivity}) and can be defined from Eq. \ref{Vout} as

\begin{equation}
S\equiv GR_f C_0\omega\gamma.
\label{sensitivity}
\end{equation}

\begin{figure}[h!]
	\centering
	\includegraphics[width=1\linewidth]{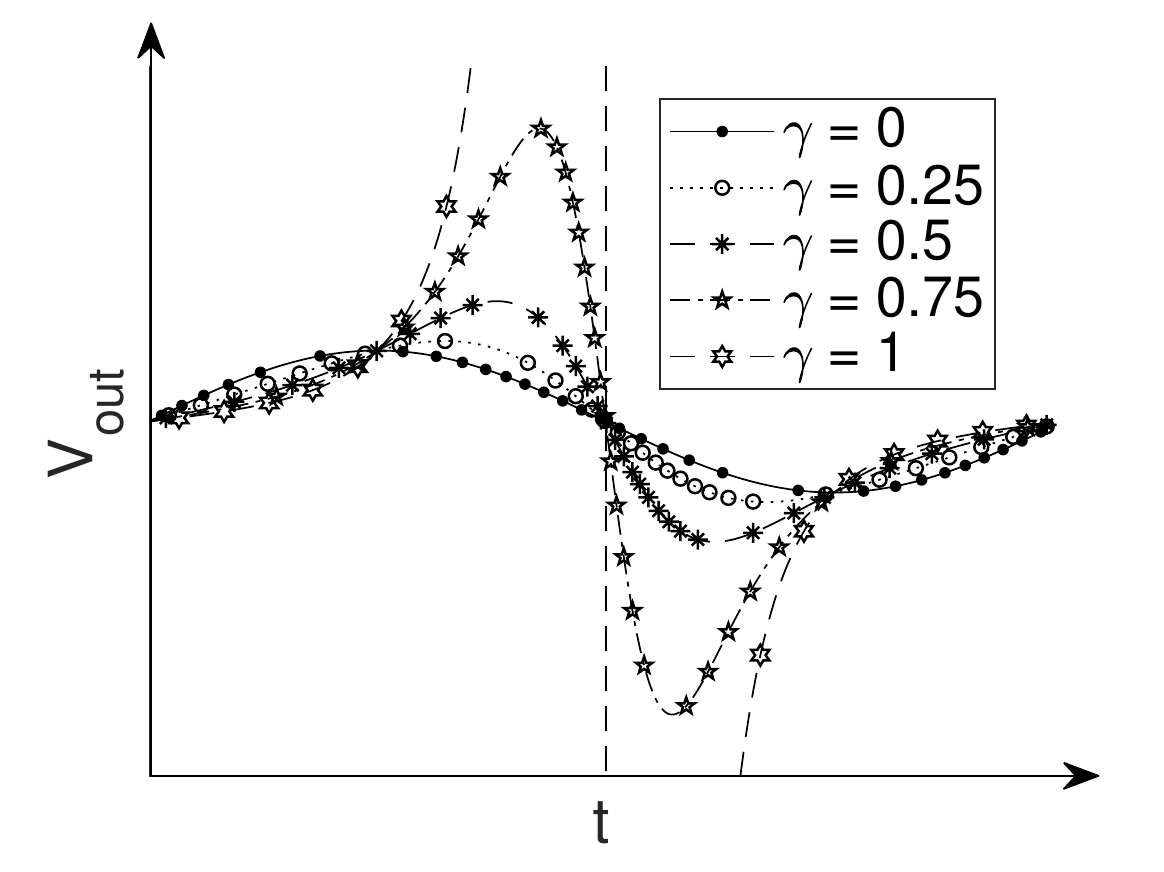}
	\caption{\label{fig:modindex}Effect of the modulation index $\gamma$ on the output signal $V_\mathrm{out}$ of a Kelvin probe circuit. For lower $\gamma$ values ($\sim\gamma<0.25$), $V_\mathrm{out}$ approaches a sine function, higher $\gamma$ values ($\sim\gamma>1$) result in unreadable $V_\mathrm{out}$ outputs, where $V_\mathrm{out}$ is characterized by large peaks over short time periods and no measurable $V_\mathrm{out}$ for the majority of the signal. Therefore, it is desirable to operate under low modulation index conditions, \textit{i.e.}, $\sim\gamma<0.25$.}
	
\end{figure}

Additionally, $V_\mathrm{out}$ is modulated by the last term of Eq. \ref{Vout}, the effect of the modulation index $\gamma$ on $V_\mathrm{out}$ can be appreciated in Fig. \ref{fig:modindex}.

If required, with the assumption of a low modulation index ($\sim\gamma<0.25$), achieved by setting a $d_1<d_0$, analyses can be simplified by reducing Eq. \ref{Vout} to

\begin{equation}
V_\mathrm{out}=GR_f C_0\omega\gamma(V_\mathrm{CPD}+V_\mathrm{b})\sin(\varphi+\alpha).
\label{VoutRed}
\end{equation}

Finally, the implementation of the off-null method for $V_\mathrm{CPD}$ determination (section \ref{offnull}) is based on the linear function $V_\textrm{out}=m_K V_\mathrm{b}+c$ originated from the relation between $V_\mathrm{b}$ and $V_\mathrm{out}$, where the Kelvin gradient $m_K$ is given by 

\begin{equation}
m_K=2GR_f C_0\omega\gamma.
\end{equation}

Since $C_0\equiv\varepsilon_r\varepsilon_0 A_p/d_0$ and $\gamma\equiv d_1/d_0$, $m_K$ can be rewritten as

\begin{equation}
m_K=\frac{2GR_f\varepsilon_r\varepsilon_0 A_p\omega d_1}{d_0^2},
\end{equation}

where it is observed that all the parameters in the numerator are constants, revealing the relation 

\begin{equation}
m_K\propto d_0^{-2},
\end{equation}

which can be exploited to implement the automatic scanning of irregular surfaces on a Kelvin probe by compensating $d_0$ variations with a corresponding $m_K$ adjust \textit{via} a feedback loop.

\section{Design considerations}

\subsection{Lateral resolution}

The lateral resolution can be defined as the ability of an imaging system to differentiate a parameter between two separate points in space. In a Kelvin probe, this property is limited mainly by the dimensions \cite{Baumgartner1988,Wicinski2016,Polyakov2010} and shape \cite{McMurray2002} of the probe tip, while also being strongly related to the mean probe-to-sample distance $d_0$. \cite{McMurray2002,Wicinski2016} High lateral resolutions are characterized by flat probe tips with small area $A_p$ and small $d_0$. \cite{Bonnet1984} In this paper we analyzed only the scenario of a flat probe tip.

\subsection{Sensitivity}

The output signal $V_\mathrm{out}$, and therefore the feasibility of the determination of $V_\mathrm{CPD}$, is influenced by the sensitivity $S$ of the Kelvin probe. Equation \ref{sensitivity} shows that reducing the tip area $A_p$, while increasing the lateral resolution, also causes a decrease in $S$. Nonetheless, is inferred that an $S$ value can be maintained as long as any reduction in $A_p$ is compensated by a proportional increase in vibration frequency $f$. Equation \ref{sensitivity} also shows that a smaller mean probe-to-sample distance $d_0$ will increase $S$. However, a smaller $d_0$ will also increase the modulation index $\gamma$, which is preferably kept low ($\sim\gamma<0.25$), the use of lower $d_0$ for increasing the sensitivity is thus limited by the relation $d_1<d_0$, necessary to guarantee a low $\gamma$ value.

\subsection{Signal-to-noise ratio}

The signal-to-noise ratio plays an important role in the accuracy of $V_\mathrm{CPD}$ determination on a Kelvin probe, for this aspect, the implementation of the off-null method might suffice. Selecting backing potential $V_\mathrm{b}$ values far enough from the null point increases the accuracy by avoiding random noise generated when $V_\mathrm{b}$ values approach the null point. \cite{Baikie1991b} Additionally, while it is possible to retrieve $V_\mathrm{CPD}$ by linear extrapolation or interpolation, the later is preferred as the former is subject to greater uncertainty. Furthermore, while any number of backing potentials can be applied, two has been found to be the optimal value that allows for the lowest uncertainty in $V_\mathrm{CPD}$ determination. \cite{Reasenberg2013}

The actuator selection and wiring of the system is also a factor to account for, \textit{e.g.}, the high voltages required to drive piezoelectric actuators can constitute a source of noise. \cite{Ritty1981} A discussion on sources of noise and troubleshooting strategies can be found in Refs. \onlinecite{Baikie1991b,Ritty1981}. 

\begin{acknowledgments}
	
The author would like to acknowledge the financial support from the Tsinghua-Berkeley Shenzhen Institute (TBSI) and the National Council of Science and Technology of Mexico (CONACyT) with the TBSI PhD Scholarship and the CONACyT Scholarship No. 424671, respectively.
	
\end{acknowledgments}

\end{document}